\documentclass[preprint]{aastex}

\newcommand{\shell}{GSH~23.0$-$0.7+117\ }
\newcommand{\shellp}{GSH~23.0$-$0.7+117.\ }
\newcommand{\shellc}{GSH~23.0$-$0.7+117,\ }

\newcommand{\kms}{\rm km\ s^{-1}}
\newcommand{\HI}{H\ {\small I}\ }

\shorttitle{GSH~23.0$-$0.7+117}
\shortauthors{Stil et~al.}

\begin{document}

\title{\shellc a neutral hydrogen shell in the inner Galaxy}

\author{J. M. Stil\altaffilmark{1}  and A. R. Taylor\altaffilmark{1}}
\author{P. G. Martin\altaffilmark{2,3} and T. A. Rothwell\altaffilmark{2}}
\author{J. M. Dickey\altaffilmark{4}}
\author{N. M. McClure-Griffiths\altaffilmark{5}}

\altaffiltext{1}{Department of Physics and Astronomy, University of Calgary, 2500 University Drive N.W., Calgary AB T2N 1N4, Canada}
\altaffiltext{2}{Department of Astronomy and Astrophysics, University of Toronto, 60 St. George Street Room 1403, Toronto ON, M5S 3H8, Canada}
\altaffiltext{3}{Canadian Institute for Theoretical Astrophysics, McLennan Labs, University of Toronto, 60 St. George Street, Toronto ON, M5S 3H8, Canada}
\altaffiltext{4}{Department of Astronomy, University of Minnesota, 116 Church Street, SE, Minneapolis, MN 55455, USA}
\altaffiltext{5}{Austalia Telescope National Facility, CSIRO, P.O. Box 76, Epping NSW 1710, Australia}

\begin{abstract}
\shell is a well-defined neutral hydrogen shell discovered in the VLA
Galactic Plane Survey (VGPS). Only the blueshifted side of the shell
was detected. The expansion velocity and systemic velocity were
determined through the systematic behavior of the \HI emission with
velocity.  The center of the shell is at
($l$,$b$,$v$)=($23\fdg05$,$-0\fdg77$,$+117\ \kms$). The angular radius
of the shell is $6\farcm8$, or 15 pc at a distance of 7.8 kpc. The \HI
mass divided by the volume of the half-shell implies an average
density $n_H = 11\ \pm 4\ \rm cm^{-3}$ for the medium in which the
shell expanded.  The estimated age of \shell is 1 Myr, with an upper
limit of 2 Myr. The modest expansion energy of $2 \times 10^{48}\ \rm
erg$ can be provided by the stellar wind of a single O4 to O8 star
over the age of the shell. The $3\ \sigma$ upper limit to the 1.4 GHz
continuum flux density ($S_{1.4} < 248\ \rm mJy$) is used to derive an
upper limit to the Lyman continuum luminosity generated inside the
shell. This upper limit implies a maximum of one O9 star (O8 to O9.5
taking into account the error in the distance) inside the \HI shell,
unless most of the incident ionizing flux leaks through the \HI
shell. To allow this, the shell should be fragmented on scales smaller
than the beam (2.3 pc). If the stellar wind bubble is not adiabatic,
or the bubble has burst (as suggested by the \HI channel maps),
agreement between the energy and ionization requirements is even less
likely.  The limit set by the non-detection in the continuum provides
a significant challenge for the interpretation of \shell as a stellar
wind bubble.  A similar analysis may be applicable to other Galactic
\HI shells that have not been detected in the continuum.

\end{abstract}

\keywords{ISM: bubbles --- ISM: kinematics and dynamics --- stars: winds, outflows --- ISM: atoms}

\section{Introduction}

The VLA Galactic Plane Survey (VGPS) is part of an international
effort to map atomic hydrogen and other tracers of the Galactic
interstellar medium with a resolution of $1'$. Previously,
large parts of the Galactic plane in the northern sky were covered by
the Canadian Galactic Plane Survey (CGPS) \citep{taylor2003}, and in
the southern sky by the Southern Galactic Plane Survey (SGPS)
\citep{mcclure2001}. The VGPS covers the first Galactic quadrant in
the vicinity of the celestial equator, for which the Very Large Array
is the most suitable instrument. The VGPS survey area extends from
Galactic longitude $18\arcdeg$ to $67\arcdeg$. The latitude coverage
varies from $|b| < 1\arcdeg$ at the low longitudes to $|b| < 2\arcdeg$
at the high longitudes.  An outline of the survey area was shown by
\citet{taylor2002}.

One important objective of these high resolution \HI surveys is to
study the effect of stellar wind and supernova explosions on the
interstellar medium. There is a rich literature on this subject, and
we limit the discussion to some examples that relate to the subject of
this paper.  The effects of stellar wind and supernovae may be
manifested on scales of hundreds of parsecs for super bubbles, chimneys
and worms, e.g.  \citet{heiles1979}, \citet{heiles1984},
\citet{normandeau1996}, \citet{mcclure2000}, \citet{english2000},
\citet{stil2001}, \citet{uyaniker2002}, \citet{mcclure2002}, to $\sim
10\ \rm pc$ for winds of single stars. Smaller bubbles originating
from a single star may be found around Wolf-Rayet stars \citep[for a
discussion of radio observations]{cappa2002} and some other early type
stars, e.g. \citet{higgs1994}, \citet{normandeau2000},
\citet{carral2002}.  An interesting question in this respect is what
fraction of the stellar wind and supernova ejecta produced in the disk
breaks out of the Galactic disk and flows into the Galactic
halo. Whether or not a breakout occurs depends on the scale of the
bubble and the scale height of \HI in the disk. In this context, small
bubbles represent events in which matter and energy ejected by massive
stars are retained in the disk. As such, smaller bubbles provide a
different perspective on the Galactic energy budget, as well as a probe of
conditions that relate to the release of enriched matter and energy
into the disk and the halo. Parsec scale \HI bubbles have become
accessible for systematic study through the recent high-resolution \HI
surveys.

The stellar winds of OB stars are driven by the ultraviolet continuum.
Therefore, a strong stellar wind and a high ionizing flux are
correlated.  A stellar wind bubble can be completely or partly ionized
due to the Lyman continuum flux of the central star. It is not clear a
priori whether the ionization of a shell can be detected in the radio
continuum images of the VGPS. Confusion with unrelated emission may
inhibit detection for larger shells. Well-documented examples of
smaller shells that would be detectable in a survey such as the VGPS
exist in the literature, e.g.  \citet{higgs1994},
\citet{cappa1999}. In this paper we present the small \HI shell \shell
discovered in the VGPS, and the implications of its non-detection in
the continuum.

\section{Observations and data reduction}
\label{data-sec}

The data presented in this paper were obtained as part of the VLA
Galactic Plane Survey (VGPS). The survey area, observing strategy and
data reduction were previously described in \citet{taylor2002}. The
VGPS survey area consists of a hexagonal grid of 990 fields observed
with the Very Large Array (VLA) D-array, separated by $25'$. Each
field was observed as a $\sim$3 minute snapshot at 2 to 5 (normally at
least 3) different hour angles.  In order to obtain sufficient
spectral resolution over the velocity range of Galactic \HI, the
3-minute observation time for each snapshot was divided into two
integrations. In each integration, left and right hand polarization
were recorded over a bandwidth of 1.56 MHz in 256 channels (6.1 kHz or
$1.28\ \kms$ per channel), but offset in frequency by 303.06 kHz (49.5
channels).  In between integrations, the frequency offset between the
polarizations was reversed. The frequency offset provides sufficient
line-free channels for continuum subtraction, and complete spectral
sampling of the \HI line over approximately $240\ \kms$. The \HI data 
presented in this paper were sampled with $0.64\ \kms$ channels.

Information on large-scale structures resolved out by the VLA mosaic
was provided by a fast survey of the VGPS area with the Robert C. Byrd
Green Bank Telescope (\HI line), and the Effelsberg 21-cm survey
\citep{reich1986,reich1990} for the continuum.

Calibration was carried out using standard procedures within AIPS. The
primary calibrators 3C286 and 3C48 were used for flux and bandpass
calibration. After calibration, the UV data were imported into MIRIAD
for further processing. At the time of submission of this paper, a
comparison of continuum source fluxes with the NVSS survey
\citep{condon1998} indicated that the VGPS flux scale is low by as
much as $30\%$ for some fields due to an increase in system
temperature by the bright \HI line emission. A field-dependent
correction based on NVSS source fluxes is currently underway. The flux
scale in this paper is based on the calibration of the GBT \HI data.
One would normally adjust the flux scale of the single dish to the
flux scale of the interferometer when the two are combined. Instead,
the flux scale of the VLA was divided by 0.85 to match the GBT
calibration in the region of the UV plane sampled by both datasets.

Continuum subtraction was done in the UV plane by fitting a linear
polynomial to the visibilities in line-free channels.  A ``dirty''
continuum-subtracted mosaic cube, fully sampled on the frequency axis,
was then created. The desired $1'$ resolution (FWHM) was obtained by
applying an appropriate Gaussian weighting function to the
visibilities.  A temporary mosaic with (projected) baselines longer
than $0.3\ \rm k\lambda$ was also created, to clean strong compact
continuum sources in absorption. These absorbed sources were
subtracted from the actual \HI images and restored with a Gaussian
beam before deconvolution of the line emission. This procedure
eliminates the side lobes around absorbed continuum sources, in
particular the radial spokes that occur on the 20\% level in the dirty
beam of VLA snapshots.  The \HI line emission was then deconvolved
with a maximum entropy algorithm \citep{sault1996,cornwell1985}.  The
deconvolved model of ``clean components'' was restored with a $1'$
circular Gaussian beam. 


\section{Results and analysis}
\label{results-sec}

Channel maps of \shell are shown in Figure~\ref{chanmap-fig}.  A ring
that is open on the lower side is best visible at velocities $113\
\kms$ to $116\ \kms$. At smaller velocities, the radius of this ring
decreases until the ring merges into a cap in the velocity range $107\
\kms$ to $110\ \kms$. The systematic behavior of the emission with
velocity is characteristic of an expanding \HI shell. The principle
characteristics that lead to this conclusion are the almost
uninterrupted ring morphology and the gradual spatial variation of the
line of sight velocity towards the cap in the center. The gradual
variation of the velocity towards the center unambiguously associates
the cap with the ring, and suggests a line of sight velocity
difference of approximately $10\ \kms$. It will be shown that this
velocity difference implies evolution on a timescale of 1 Myr, which
indicates that \shell is a transient phenomenon in the interstellar
medium.

The circular shape and the regular behavior of the emission as a
function of velocity indicate a high degree of symmetry. However, two
significant asymmetries are observed. First, virtually no emission of
the shell was detected at velocities beyond $\sim 120\ \rm km\
s^{-1}$. Instead, the ring-like structure in the channel maps breaks
up and fades into the noise at this velocity. Inspection of \HI line
profiles near the center of the shell shows no evidence for a
redshifted cap to the level of the noise. This implies that the
blueshifted side of the shell is at least 10 to 20 times brighter than
the redshifted side. The \HI column density of the blueshifted cap in
the center of the shell is $2 \times 10^{20}\ \rm cm^{-2}$.  The
second asymmetry is that the lower side of \shell also appears open.
This opening is flanked on both sides by filaments best visible in the
channels around $113\ \rm km\ s^{-1}$.

The one-sided \HI structure could at first sight also be interpreted
as a bow shock. The \HI brightness in Figure~\ref{chanmap-fig} is
fairly symmetric relative to the axis $l=23\fdg05$. If \HI brightness
is primarily a measure of the length of the line of sight through the
structure, the line of sight through the top is much longer than
through the bottom. This can be the result of a conical geometry, in
particular a bow shock due to a moving source with its velocity along
the axis of symmetry.  To explain the observed brightness
distribution, the line of sight through the top of \shell should be
approximately tangential to the surface of the bow shock.  The
velocity of the source with respect to the ambient medium would have a
component towards the observer (blueshifted cap), and a component
towards the Galactic plane. No redshifted side is expected for a bow
shock, because the bow shock is open on the far side. However, a bow
shock geometry was rejected because of the implied velocity
structure. The expansion velocity of a bow shock is locally
perpendicular to the surface of the shock. If the line of sight
through the top of \shell is tangential to the surface of the bow
shock, the expansion velocity is perpendicular to the line of sight
only at this location. For other lines of sight, the expansion
velocity has a line of sight component in the direction of the
observer.  This implies that the line of sight through the top of
\shell should dominate the structure in Figure~\ref{chanmap-fig} at
the most positive velocities.  The brightness distribution in
Figure~\ref{chanmap-fig} does not show the expected concentration of
emission towards the top of \shell for velocities $v > 115\ \kms$.

Since only the approaching side of the shell was detected, a
determination of the expansion velocity and the central velocity by
visual inspection of the channel maps is not feasible. Following
\citet{stil2001}, a thin shell model was fitted to the observed line
of sight components of the expansion velocity.  We fitted Gaussians to
the observed line profile at each position to determine the line of
sight component of the expansion velocity.  On the lower side of the
shell the low signal to noise ratio often inhibited a determination of
the line of sight velocity.  After a number of preliminary fits of the
thin shell model, measurements located within 6.6 arcminutes from the
center of the shell were included in the analysis.

\begin{figure}
\epsscale{1}
\plotone{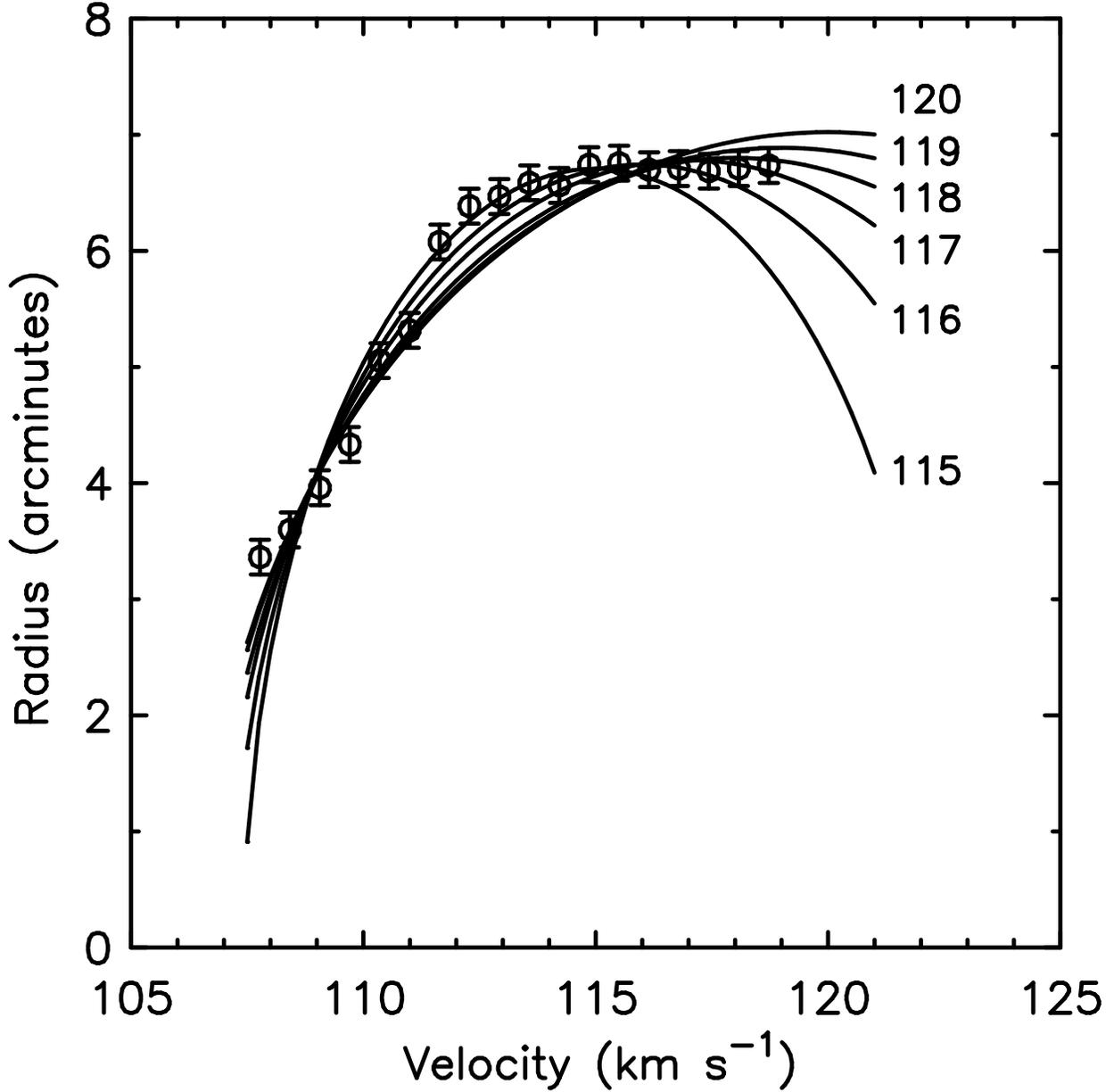}
\caption{
Observed radii of the ring of emission associated with the shell as a
function of velocity. Error bars indicate the radial bin size used for
azimuthal averaging.  The curves represent thin expanding shell
models that best fit the observed line of sight velocities for a fixed
central velocity. The central velocity in $\rm km\ s^{-1}$ for each
model is indicated.
\label{SBfit-fig}
}
\end{figure}

Models with a systemic velocity less than $117\ \rm km\ s^{-1}$ resulted
in a larger value of $\chi^2$ and were rejected.  Models with
a systemic velocity larger than $117\ \rm km\ s^{-1}$ fit the observed
line of sight velocities equally well. In this case, the effect of a
larger systemic velocity is compensated by a correspondingly larger
expansion velocity and a larger radius. This degeneracy is mainly due
to the absence of data for the redshifted side of the shell.  The
degeneracy between the systemic velocity and the expansion velocity was
resolved by considering the variation of the radius of the ring of
emission in the channel maps with velocity.  Figure~\ref{SBfit-fig}
shows the distance of the peak of the azimuthally averaged brightness
profile as a function of velocity. The curves represent restricted
thin-shell model fits in which the systemic velocity was fixed to the
values indicated in Figure~\ref{SBfit-fig}. The data are best
represented by the curve corresponding with a systemic velocity of
$117\ \rm km\ s^{-1}$, with an uncertainty of about $1\ \rm km\
s^{-1}$. The scatter at smaller radii/velocities is a result of the
fact that no clear ring is visible in the corresponding channels. The
model with systemic velocity $117\ \kms$ has an expansion velocity
$v_e = 9\ \pm\ 1\ \rm km\ s^{-1}$, where the main source of
uncertainty is the value of the systemic velocity. The properties of
\shell are summarized in Table~\ref{shellpar-tab}.

\begin{figure}
\plotone{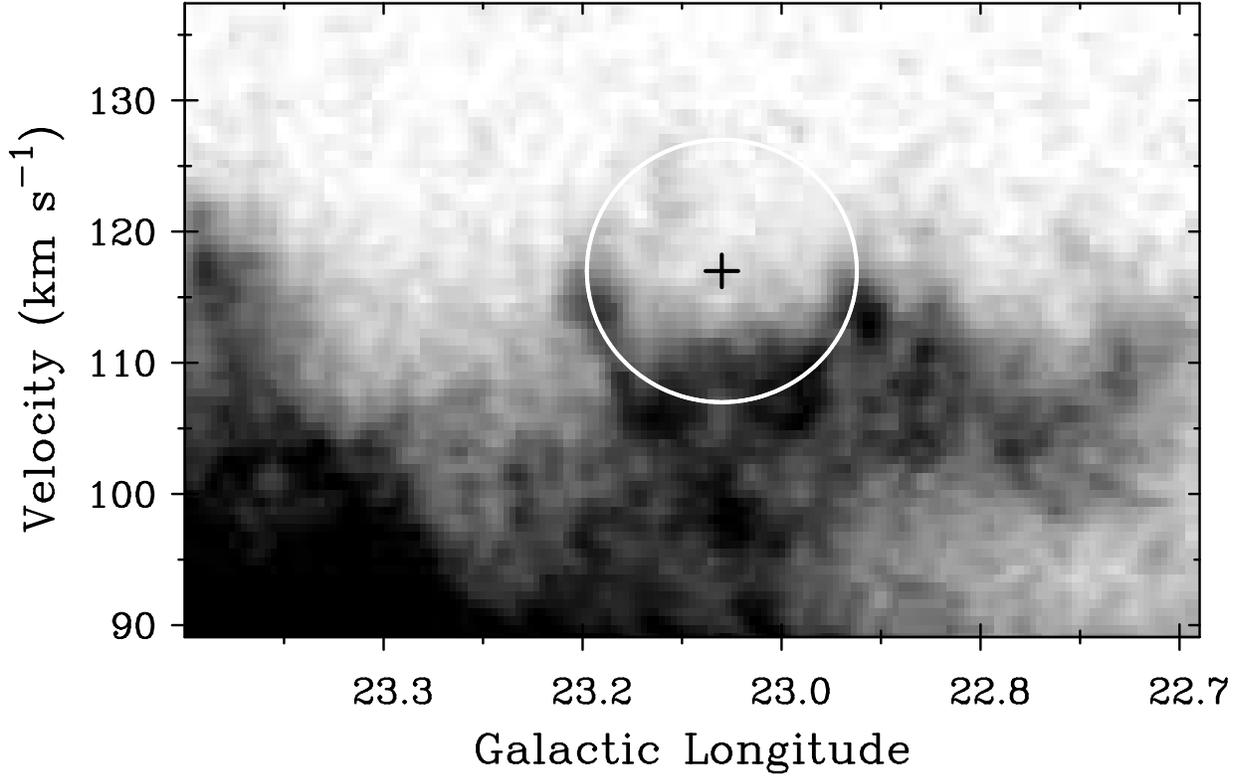}
\caption{
Longitude-velocity map through the center of the shell. Brightness
temperature is shown as gray scales linearly from 0 K (white) to 50 K
(black). The ellipse indicates the best-fit model with parameters
listed in Table 1.
\label{XVmap-fig}
}
\end{figure}

Figure~\ref{XVmap-fig} shows the position-velocity diagram through the
center of the shell at constant Galactic latitude. The best-fit model
is shown as a white ellipse.  Emission at $100\ \rm km\ s^{-1}$ is
clearly identified as a separate velocity component in
Figure~\ref{XVmap-fig}.

The central velocity of \shell is close to the extreme positive
velocity of Galactic \HI at longitude $23\arcdeg$, so the distance is
relatively well determined.  Assuming a flat rotation curve with
$R_0=8.5\ \rm kpc$ and $V_0=220\ \rm km\ s^{-1}$, the Galactocentric
distance is 3.6 kpc, and the heliocentric distance is 6.6 or 9
kpc. This location is close to the extreme end of the Galactic bar
\citep{cole2002}.  The assumption of circular orbits in the derived
kinematic distance may introduce a significant error.  Therefore, we
consider the difference between the near and far point not significant
and adopt a distance $7.8\ \pm\ 2\ \rm kpc$. At this distance, the
radius of \shell is $15\ \pm\ 4\ \rm pc$, and it is located at $110\ \pm\ 30\
\rm pc$ below the Galactic plane.

The 21-cm line flux in the velocity range $105\ \rm km\ s^{-1}$ to
$125\ \rm km\ s^{-1}$ in a circular aperture enclosing the shell is
$131\ \rm Jy\ km\ s^{-1}$.  Variation of the velocity range by $2\ \rm
km\ s^{-1}$ on each side changes the flux by about 20\%, mainly due to
a confusing component at $100\ \kms$.  The \HI mass of the shell is
then $M_{\rm HI} = 1.9 \times 10^{3}\ \rm M_\odot$.  The total mass of
the shell, including 30\% helium, is $M_{\rm S}=2.5 \times 10^{3}\ \rm
M_\odot$. The error in the mass is $\sim 50\%$, dominated by the error
in the distance. The kinetic energy associated with the expansion is
$E_{\rm kin}=2 \times 10^{48}\ \rm erg$, and the expansion momentum
(the scalar product of mass and expansion velocity) is $2 \times
10^{4}\ \rm M_\odot\ \kms$. The mass of the shell divided by the
volume of the shell provides an estimate of the average density of the
medium in which the shell expanded. The resulting density scales only
inversely proportional with the assumed distance, and relies further
on the plausible assumption that the mass of the shell is dominated by
swept-up interstellar gas.  The mass of the observed half shell
suggests a mean density $n_{\rm H}=11\ \pm\ 4\ \rm cm^{-3}$ when
distributed uniformly over the volume of the half-sphere. The error
estimate incorporates the effect of uncertainties in the HI line flux,
the distance, and the angular radius of the shell, added in
quadrature.  This density seems quite high, especially for a distance of
110 pc from the Galactic mid plane.

The kinematic age of the shell assuming an expansion law $R \sim
t^{0.6}$, appropriate for adiabatic stellar wind bubbles
\citep{weaver1977}, is $t = 0.6 R/v_{\rm e}=1.0\ \rm Myr$. An upper
limit to the age of the shell is $t_{\rm max}=R/v_{\rm e}=1.6\ \rm
Myr$. Taking into account the uncertainties in the distance and the
expansion velocity, it is unlikely that the age of the shell is more
than 2 Myr. 

\begin{figure}
\plotone{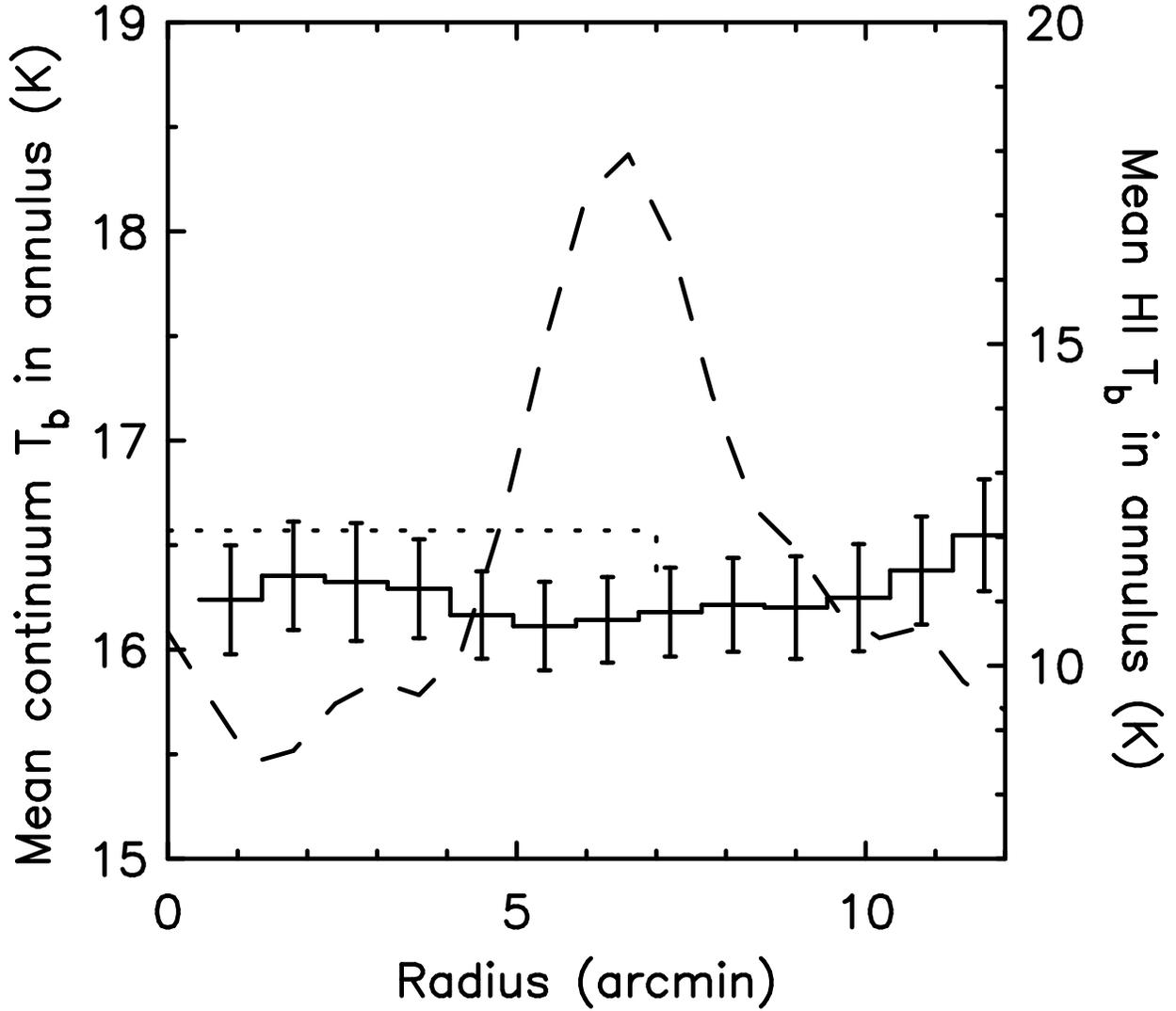}
\caption{ Azimuthally averaged surface brightness profile of continuum
brightness (histogram; scale on the left axis) and \HI at $116.7\
\kms$ (dashed curve; scale on the right axis). The peak of the dashed
line marks the radius of the \HI shell.  The error bars indicate the
error in the mean intensity, defined as the r.m.s. intensity
fluctuations divided by the square root of the number of independent
beams per annulus.  The dotted line shows the surface brightness that
corresponds to the $3\sigma$ upper limit to the radio continuum flux
density $S_\nu < 248\ \rm mJy$ derived in the text.
\label{cont-radprof-fig}
}
\end{figure}

Figure~\ref{cont-fig} shows the continuum emission in the vicinity of
\shellp There is no visible counterpart of the \HI shell in the
continuum image. The faint filamentary structure at longitude
$23\arcdeg$ extends at least $0\fdg3$ further towards the Galactic
equator, and is probably unrelated to \shellp
Figure~\ref{cont-radprof-fig} shows the azimuthally averaged surface
brightness in the continuum image as a function of distance from the
center of the \HI shell. The error bars in
Figure~\ref{cont-radprof-fig} (typically 0.24 K) indicate the
r.m.s. intensity, divided by the square root of the number of
independent beams in each annulus. This is the error in the mean
intensity per annulus, taking into account the variance of the
background.  The error bars do not decrease with radius because the
r.m.s. intensity increases with radius due to variation of the
background. Figure~\ref{cont-radprof-fig} shows significant excess
above the continuum background associated with the \HI shell.  The $3
\sigma$ upper limit to the continuum brightness of each annulus above
the Galactic background is $T_{\rm b,shell} < 0.72\ \rm K$. The annuli
contribute to the flux of the shell with weight proportional to the
radius of the annulus. The variance of the weighted sum of the
independent annuli is a weighted sum of the variances, where the
weights are proportional to the square of the radius of each
annulus. The corresponding $3 \sigma$ upper limit to the flux density
of \shell in a circular aperture with radius $7'$ is $S_\nu < 248\ \rm
mJy$. The brightness profile of a disk of uniform brightness with this
flux density is shown as a dotted line in
Figure~\ref{cont-radprof-fig}. A stronger upper limit may be found if
the continuum emission is assumed to be in a limb-brightened shell
similar to the observed \HI shell. This would be the case if an
ionization front is trapped by the \HI shell.  Adding the
contributions of annuli with radii between $4'$ and $7'$ results in a
$3 \sigma$ upper limit $S_\nu < 221\ \rm mJy$. 

The upper limit to the continuum brightness provides a limit to the
emission measure for free-free emission through the relation $\int
n_i^2 dl = 562\ T_b\ \rm cm^{-6}\ pc$ (for electron temperature $T_e =
7000\ \rm K$). For the limb brightened edge of a shell, the length of
the line of sight is of the order of the radius of the shell. The
upper limit $T_b < 0.72\ \rm K$ thus implies that the r.m.s. ionized
density is bounded by $<n_i^2>^{0.5}\ < 5\ \rm cm^{-3}$.  The upper
limit $S_\nu < 248\ \rm mJy$ implies the ionized mass is bounded by
$M_i < 1.5 \times 10^3\ \rm M_\odot$ \citep{mezger1967}. The upper
limit $S_\nu < 221\ \rm mJy$ for a shell morphology corresponds with
$M_i < 1.3 \times 10^3\ \rm M_\odot$. These upper limits are smaller
than the \HI mass of \shellc which shows that a completely ionized
bubble with the same mass as \shell would have been detected in the
VGPS continuum image.
 
The foreground extinction was estimated from the total hydrogen column
(atomic and molecular at positive velocities) in the direction of
\shellc assuming half of the hydrogen column is in the foreground,
$N_{\rm H}/E(B-V) = 5.8 \times 10^{21}\ \rm cm^{-2}\ mag^{-1}$
\citep{bohlin1978}, and $A_V/E(B-V) = 3.1$. The resulting foreground
extinction is $A_V = 8.6$ magnitudes. As the line of sight to \shell
is 110 pc from the mid plane at 7.8 kpc, the assumption that half the
hydrogen column is in front of \shell leads to a probable
underestimate of the foreground extinction.  No counterpart of \shell
was found in the VTSS H$\alpha$ survey \citep{dennison1999}, probably
due to the high foreground extinction.  Also, no trace of the shell
was found in the IRAS $60\mu$ and $100\mu$ maps or in the MSX survey
\citep{price2001}. The only entry in the SIMBAD catalog for a position
inside the \HI shell is the IRAS source IRAS 18332-0907
($l$,$b$)=($23.1$,$-0.74$). The spectrum of this source increases with
wavelength, excluding the possibility that it is an obscured star or
cluster of stars.

There is no CO emission directly associated with the shell in the
survey of \citet{dame2001}. However, a molecular cloud was found at
($l$,$b$,$v$)=($22.78$,$-1.25$,$115.7$).  The projected distance of
this cloud to the center of \shell is 68 pc.

\section{Discussion}
\label{discussion-sec}

\shell is a thin neutral gas shell expanding with a velocity of 9
$\kms$.  Such shells are believed to form when supernovae and or
stellar winds interact with the interstellar medium. Although the
source is most likely isotropic, the \HI morphology of \shell is
clearly one-sided. It is unlikely that the part of the shell not seen
in \HI is in fact ionized. If the mass surface density of \shell is to
be uniform, the mass of the ionized part of the shell must be similar
to or more than the \HI mass of the neutral part, which is
inconsistent with the upper limits to the continuum emission of the
shell. It is therefore more likely that the surface density of \shell
is not uniform.  The structure in the channel maps
(Figure~\ref{chanmap-fig}) suggests that the bubble may have burst and
that the hot gas interior to the HI shell might flow out of the
bubble. The open side of the shell is directed away from the Galactic
equator, and away from the observer, suggestive of a density gradient
in these directions. The high average density of the ambient medium
$n_H = 11\ \pm\ 4\ \rm cm^{-3}$ found in Section~\ref{results-sec} is
consistent with the presence of a relatively dense cloud in the
vicinity of the source of the shell.  A similar front-to-back
asymmetry has been observed in other \HI shells, e.g.
\citet{vdwerf1990}, \citet{cappa1999}, \citet{stil2001}. A partly
ionized shell was not excluded by \citet{vdwerf1990}, but a density
gradient in the surrounding medium was also suggested by the presence
of molecular gas on the side of the \HI half shell. \citet{cappa1999}
interpreted the \HI filament wrapped around the Wolf-Rayet ring nebula
NGC~2359 as the stellar wind bubble running into a denser cloud on one
side. Most of the $\sim 320\ \rm M_\odot$ neutral mass was reported to
be in a hemispherical shell expanding at 6-7 $\kms$.

A first indication of what could be the source of \shell is the
expansion kinetic energy of $2 \times 10^{48}\ \rm erg$. This amount
of energy is readily available in the wind of a single massive star,
and only $\sim 0.1\%$ of the kinetic energy released in a single
supernova explosion. We shall therefore consider the stellar wind of a
single star or a few stars as the energy source of \shellp

The star(s) inside the shell should be able to provide the expansion
energy of the shell.  First we determine the average mass loss rate
over the age of the shell from the expansion energy.  This mass loss
rate is then converted into an equivalent number of OB stars.  The
wind luminosity $L_w = {1 \over 2} \dot M v_w^2$, for a stellar mass
loss rate $\dot M$ and wind velocity $v_w$ follows from the energy
equation
$$
\dot M v_w^2 t = { M_S v_e^2 \over  f_w \epsilon_S},  \eqno (1)
$$
with $f_w$ the fraction of the sphere covered by the \HI shell,
$\epsilon_S$ the efficiency for the conversion of wind kinetic energy
into kinetic energy of the shell \citep{treffers1982}, $t$ the age of
the shell, $M_S$ the mass of the shell and $v_e$ the expansion
velocity of the shell. The factor $f_w$ takes into account that the
expansion energy of \shell incorporates only the incomplete \HI shell,
whereas the source of the stellar wind is isotropic.

Inspection of the \HI channel maps (Figure~\ref{chanmap-fig}) suggests
that $f_w \approx 0.33$. The efficiency $\epsilon_S = 0.2$ for an
adiabatic stellar wind bubble was adopted \citep{weaver1977}.  This
efficiency may be much lower \citep[for a discussion]{vburen1986}, but
it is unlikely to be much higher. The consequences of $\epsilon_S <
0.2$ will be discussed later in relation to the effect of other model
parameters.

The wind velocity of OB stars ranges from 1000 $\kms$ to 3000
$\kms$. Accepting a factor 2 uncertainty in the resulting mass loss
rate we assume a wind velocity of $2000\ \kms$. The average mass loss
rate over the lifetime of the shell is then $\rm \dot M = 4 \times
10^{-7}\ M_\odot\ yr^{-1}$.  The average mass loss rate was converted
into an equivalent number of OB stars with the empirical relation
between mass loss rate and luminosity ${\rm log} (\dot M) = 1.69\ {\rm
log} (L) - 15.4$, with $\dot M$ in $\rm M_\odot\ yr^{-1}$ and $L$ in
$L_\odot$ \citep{howarth1989}. The equivalent number of OB stars is
the total luminosity derived from this relation, divided by the
luminosity of a single star of a specific spectral type.  Stellar
properties as a function of spectral type were taken from
\citet{vacca1996}. This conversion introduces an additional
uncertainty of a factor 2 due to the scatter in the empirical relation
between mass loss rate and luminosity.

The expansion momentum of the shell provides a consistency check for
the mass loss rate derived in this way.  The ratio $\pi_S$ of the
momentum of the shell to the momentum carried by the wind over the age
of the shell ($\dot M v_w t$), is more than 1 for an adiabatic
bubble. This is possible because the expansion momentum, the
scalar product of the mass of the bubble and its expansion velocity,
is increased by the pressure of the hot interior of the bubble. The
derived mass loss rate implies that the total wind momentum over the
age of \shell is $800\ \rm M_\odot\ \kms$, so that $\pi_S=25$. This
ratio is within the range found by \citet{vburen1986} for nine stellar
wind bubbles.

\begin{figure}
\plotone{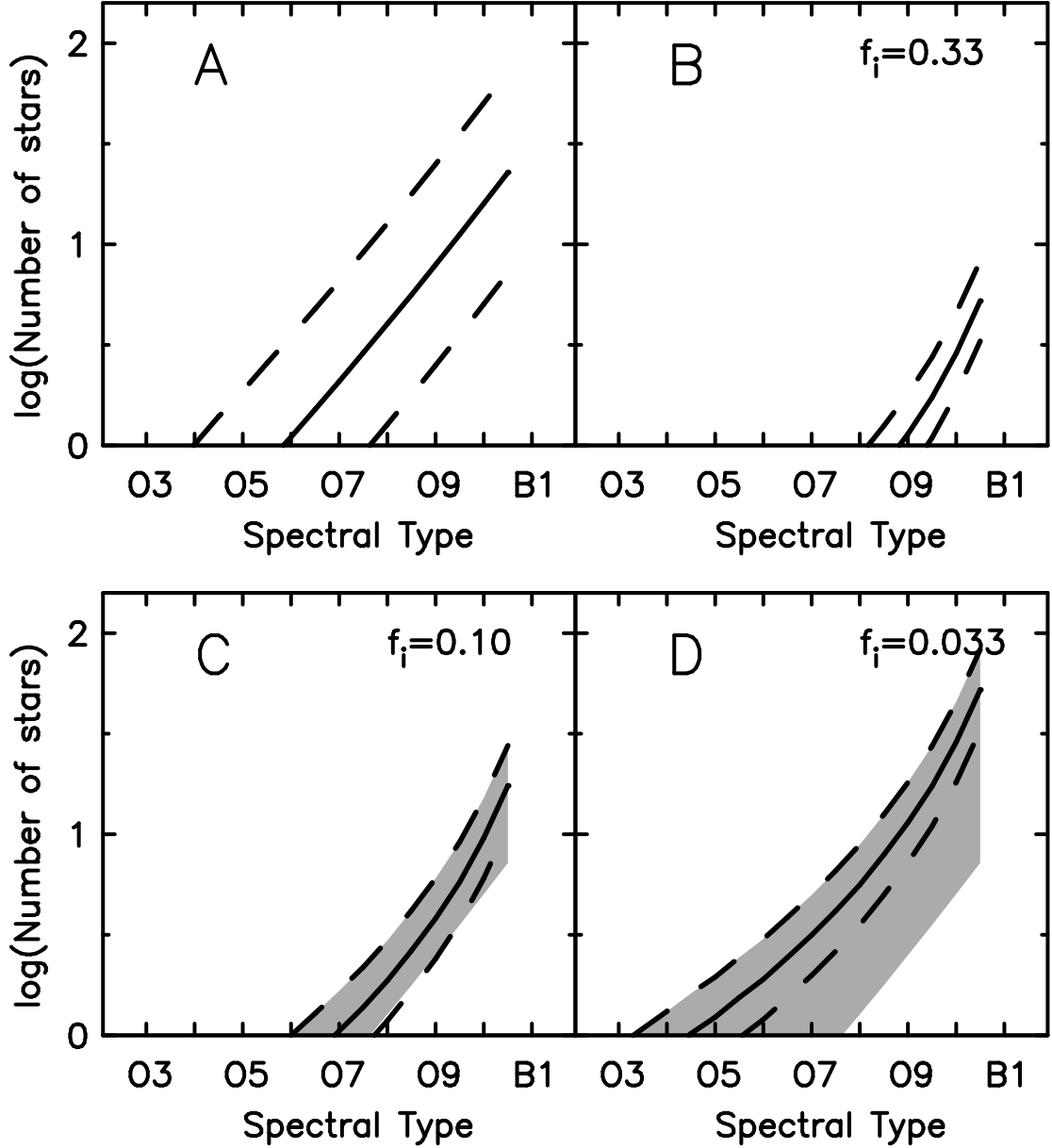}
\caption{ A; Number of stars as a function of spectral type that is
required to provide the expansion energy of the shell. B: Number of
stars inside the shell that is consistent with the upper limit to the
continuum for $f_i = 0.33$. C: As B, but for $f_i = 0.1$. D: As B, but
for $f_i = 0.033$. The gray shaded areas indicate where the
constraints of energy and ionization are both satisfied.
\label{stellar-fig}
}
\end{figure}

Figure~\ref{stellar-fig}~A shows the equivalent number of OB stars
necessary to provide the expansion energy of \shell through stellar
winds as a function of spectral type. The dashed lines indicate the
uncertainty assuming a factor 2 uncertainty in the expansion energy, a
factor 2 due to the assumption of a constant wind velocity, and a
factor 2 scatter in the empirical relation between luminosity and mass
loss rate, added in quadrature. Figure~\ref{stellar-fig}~A indicates
that the expansion energy of the shell is consistent with (the
equivalent of) a single star of spectral type O4 to O8.

If \shell is a stellar wind bubble, it can be expected that the
star(s) that provided the stellar wind are still present within the
shell because the age of the shell is less than the main sequence
lifetime of the most massive stars. Although the following discussion
is based on the initial assumption that OB stars still exist inside
the \HI shell, the results that follow can in principle suggest
otherwise.  The ionizing flux of these stars is bounded by the upper
limit of the radio continuum flux derived in Section~\ref{results-sec}
through the relation \citep{rubin1968}
$$
f_i L_c = 4.76 \times 10^{39}\ \nu_{\rm GHz}^{0.1}\ d_{\rm pc}^2\ S_{\nu;\rm mJy}\ T_e^{-0.45}. \eqno (2)
$$
Absorption of Lyman continuum photons by dust is neglected.  In
equation (2) it is assumed that a fraction $(1-f_i)$ of the Lyman
continuum photons escapes from the bubble. This may be because the
shell is incomplete, or because the low density part of the shell is
fully ionized and does not trap the ionization front.  As no evidence
for an ionized part of the shell was detected, it can be assumed that
$f_i = f_w = 0.33$, provided the observed \HI shell intercepts all of
the incident ionizing photons. If ionizing radiation leaks through
holes in the \HI shell not visible at the resolution of the current
data, $f_i < 0.33$. Stellar wind bubbles are subject to a number of
hydrodynamic instabilities that can cause such fragmentation of the
shell \citep{rozyczka1985,garcia1995,garcia1996,freyer2003}.
Inserting $\nu = 1.42\ \rm GHz$, $d=7.8\ \rm kpc$, $S_{\nu} < 248\ \rm
mJy$, and electron temperature $T_e = 7000\ \rm K$, we have ${\rm
log}(f_i L_c)< 48.5$.

The number of stars of each spectral type that is consistent with the
upper limit to the ionizing flux that can be generated inside the \HI
shell is shown in Figure~\ref{stellar-fig}~B-D for three values of
$f_i$. Stellar parameters were taken from \citet{vacca1996}.  Note
that the solid lines in Figure~\ref{stellar-fig}~B-D represent {\it
upper limits} derived from the upper limit to the continuum flux.  The
dashed lines in Figure~\ref{stellar-fig}~B-D indicate the effect of
the error in the distance on this upper limit. The shaded areas in
Figure~\ref{stellar-fig}~B-D indicate the areas where both the energy
and the ionization constraints are satisfied, taking into account the
uncertainties indicated by the dashed lines.  No consistent solution
was found for $f_i=0.33$ (Figure~\ref{stellar-fig}~B). This is the
case where all the ionizing radiation emitted by the central star(s)
in the direction of the observed \HI shell is trapped by the shell.
Some overlap between the energy and ionization constraints is found
for $f_i = 0.1$ (Figure~\ref{stellar-fig}~C). In this case,
two-thirds of the ionizing flux emitted in the direction of the \HI
shell by the central star(s) escapes through holes in the \HI shell
not visible in the present data. In Figure~\ref{stellar-fig}~D, 90\%
of the ionizing radiation escapes in this way. However, this high
porosity of the \HI shell is less plausible. The detected part of the
\HI shell appears continuous in Figure~\ref{chanmap-fig}. Therefore, a
high porosity implies that the shell must be highly fragmented on
scales smaller than the beam (2.3 pc).

The size of the region in which both constraints are satisfied,
depends also on the model parameters $f_w$ and $\epsilon_S$.  The
value of $f_w$ is relatively well constrained from the morphology in
the \HI channel maps. An estimate of the parameter $\epsilon_S$, the
ratio of the expansion kinetic energy of the shell to the kinetic
energy deposited in the form of stellar wind, involves detailed and in
some cases poorly understood physics of the evolution of stellar wind
bubbles. However, only an upper boundary to the value of $\epsilon_S$
is required here. The value $\epsilon_S = 0.2$ derived by
\citet{weaver1977} for the early stage of evolution (time scale of a
few thousand years) provides an adequate limiting value for the purpose
of this paper. The conclusions do not change qualitatively unless
$\epsilon_S > 0.4$, which is an unlikely high efficiency. A stellar
wind bubble in the momentum-conserving phase would have $\epsilon_S
\ll 1$.  Also, if the bubble did burst, the internal pressure would
drop and $\epsilon_S \ll 1$ should be adopted. This eliminates the
small overlap between the energy requirement and the ionization
constraint in Figure~\ref{stellar-fig}, which assumes
$\epsilon_S=0.2$.  Therefore, the case $\epsilon_S \ll 1$ creates an
even bigger challenge to reconcile the expansion energy of the shell
with the limit to the ionizing flux that can be generated inside the
shell.

It is concluded from Figure~\ref{stellar-fig} that the equivalent of a
single O4 to O8 star is required to provide the expansion energy of
\shell through a stellar wind. However, the upper limit to the
continuum flux of the shell excludes stars with spectral type earlier
than O8 unless a significant part of the ionizing flux escapes through
holes in the \HI shell. A small value of $f_i$ is not implied by the
present data, but higher resolution \HI observations could verify
this.  If the bubble burst, as suggested by the morphology in the \HI
channel maps, or if the bubble is in the momentum conserving phase,
the small values of $f_i$ adopted in Figure~\ref{stellar-fig}~C and D
also fail to provide a solution that satisfies both constraints.  It
is more likely that a contradiction exists between the spectral type
of stars deduced from the dynamics of the shell and the spectral types
deduced from the upper limit to the radio continuum flux. In view of
this contradiction, the initial assumption that \shell is a stellar
wind bubble may have to be reconsidered.

If \shell is not a stellar wind bubble, its energy source may still be
of stellar origin. The \HI shell itself provides little or no clue to
the nature of its source other than its global properties such as the
expansion energy.  The shell is as much shaped by the local
interstellar medium as it is by its source.  We note that the modest
expansion energy of \shell can be provided by a single star through
stellar wind or through a supernova explosion. A supernova explosion
would provide the energy without leaving an ionizing source in the
shell. One may expect to see a non-thermal radio continuum supernova
remnant in this case, although it is not clear whether a radio
supernova remnant should still be detectable at the age of \shell.
More sensitive observations of possible ionized gas associated with
\shell, and a similar analysis of other \HI shells with or without
continuum counterparts can help clarify this issue.


\section{Conclusions}

We report the discovery of the small \HI shell \shell in the VLA
Galactic Plane Survey (VGPS). The physical parameters of \shell are
well constrained because its velocity places it close to the tangent
point.  At $(l,b,v) =$ ($23\fdg05$,$-0\fdg77$,$+117\ \rm \kms$), the
distance of the shell is $7.8\ \pm\ 2\ \rm kpc$. The expansion kinetic
energy is found to be $2 \times 10^{48}\ \rm erg$ for a shell mass of
$2.5\times 10^3\ \rm M_{\odot}$ and expansion velocity $9\ \pm\ 1\
\kms$. The age of the shell is $1\ \rm Myr$ with a strong upper limit
of $2\ \rm Myr$. The average density of the medium in which the shell
expanded derived from the mass and the volume of the shell is $n_H =
11\ \pm 4\ \rm cm^{-3}$. This relatively high density and the
one-sided morphology of the \HI shell are suggestive of a density
gradient or a cloud in the vicinity of the shell.  The \HI shell has
no counterpart in the VGPS 1.4 GHz continuum image, with an upper
limit to the 1.4 GHz flux density $S_{1.4} < 248\ \rm mJy$.

The interpretation of \shell as a stellar wind bubble is tested by
combining the energy requirements of the shell with a limit to the
number of OB stars inside the shell derived from the upper limit to
the 1.4 GHz continuum emission.  If \shell is an adiabatic stellar
wind bubble \citep{weaver1977}, the expansion energy requires the
equivalent of the stellar wind of a single O4 to O8 star.  However,
the $3\sigma$ upper limit to the 1.4 GHz continuum flux density
excludes more than one O9 star (O8 to O9.5 given the uncertainty in
the distance), unless most of the incident ionizing flux leaks through
the \HI shell. For this to be the case, the \HI shell should be highly
fragmented on scales smaller than the beam (2.3 pc). The energy
requirements for the shell are significantly larger if the bubble is
not adiabatic, or if the bubble has burst, as suggested by the
morphology in the \HI channel maps. In this case, the discrepancy
between the energy requirements of the shell and the maximum number of
OB stars allowed inside the shell is even larger.

We conclude that the interpretation of the low-energy \HI shell \shell
as a stellar wind bubble is questionable in view of the absence of
continuum emission associated with the shell. A similar argument may
be applicable to other Galactic \HI shells that have not been detected
in the continuum.

\section{Acknowledgements}

The National Radio Astronomy Observatory is a facility of the National
Science Foundation operated under cooperative agreement by Associated
Universities, Inc. This research has made use of the SIMBAD database,
operated at CDS, Strasbourg, France.  The VGPS is supported by a grant
to ART from the Natural Sciences and Engineering Council of Canada.
We thank the anonymous referee for useful comments about the manuscript.

\clearpage

\begin{deluxetable}{lr} 
\tablecolumns{2}
\tablewidth{0pc} 
\tablecaption{Observed and derived properties of the shell} 
\tablehead{ 
\colhead{Quantity } &  Value \ \ \ \  
}   
\startdata
 Center longitude    &  23\fdg050 $\pm$ 0\fdg003     \\
 Center latitude     & -0\fdg774  $\pm$ 0\fdg003     \\
 Systemic velocity   &      117  $\pm$ 1         $\rm km\ s^{-1}$ \\
 Expansion velocity  &        9  $\pm$ 1         $\rm km\ s^{-1}$ \\
 Kinematic distance  &      7.8  $\pm$ 2         kpc \\
 Angular radius      &      6\farcm8  $\pm$ 0\farcm4  \\
 Radius              &       15  $\pm$ 4         pc \\
 \HI mass             &      1.9                 $\times 10^3\ \rm M_\odot$ \\ 
 Neutral gas mass    &      2.5                 $\times 10^3\ \rm M_\odot$ \\
 Expansion energy    &        2                 $\times 10^{48}\ \rm erg$ \\
 Expansion momentum  &        2                 $\times 10^{4}\ \rm M_\odot\ \kms$ \\
 $T_{b, 1.4\ \rm GHz}$    &  $< 0.72$                  K \\
 $S_{1.4\ \rm GHz}$       &  $< 248$                  mJy     \\
\enddata
\label{shellpar-tab}
\end{deluxetable} 

\clearpage

\begin{figure}
\epsscale{0.8}
\caption{ Channel maps of \shellp The velocities of each channel is
indicated in the lower right corner. Brightness temperature is
indicated in gray scales linearly from 0 K (white) to 60 K
(black). The white contour indicates the 40 K level.
\label{chanmap-fig}
}
\end{figure}

\begin{figure}
\caption{ Continuum emission in the direction of \shellp Gray scales
indicate continuum brightness temperature linearly from 12 K to 40
K. White contours indicate continuum levels from 30 K to 150 K at
intervals of 15 K. The black contour shows the outline of \shell at
the the 22 K level at velocity $114.2\ \kms$. The $60''$
beamsize is shown in the lower right corner.
\label{cont-fig}
}
\end{figure}

\begin{thebibliography}{}

\bibitem[Bohlin et~al. (1978)]{bohlin1978} Bohlin, R. C., Savage, \& B. D., Drake, J. F. 1978, \apj, 224, 132
\bibitem[Cappa et~al. (1999)]{cappa1999} Cappa, C. E., Goss, W. M., Niemela, V. S., \& Ostrov, P. G. 1999, \aj, 118, 948
\bibitem[Cappa et~al. (2002)]{cappa2002} Cappa, C. E., Goss, W. M., \& Pineault, S. 2002, \aj, 123, 3348
\bibitem[Carral et~al. (2002)]{carral2002} Carral, P., Kurtz, S. E., Rodr\'iguez. L. F., Menten, K. F., Cant\'o, G., \& Arceo, R. 2002, \aj, 123, 2574
\bibitem[Cole \& Weinberg (2002)]{cole2002} Cole, A. A., \& Weinberg, M. D. 2002, \apj, 574, L43
\bibitem[Condon et~al. (1998)]{condon1998} Condon, J. J., Cotton, W. D., Greisen, E. W., Yin, Q. F., Perley, R. A., Taylor, G. B., \& Broderick, J. J. 1998, \aj, 115, 1693
\bibitem[Cornwell \& Evans (1985)]{cornwell1985} Cornwell, T. J., \& Evans, K. F. 1985, \aap, 143, 77
\bibitem[Dame et~al. (2001)]{dame2001} Dame, T. M., Hartmann, D., \& Thaddeus, P. 2001, \apj 547, 792
\bibitem[Dennison et~al. (1999)]{dennison1999} Dennison, B., Simonetti, J. H., \& Topasna, G. A. 1999 \baas, 195, 5309
\bibitem[English et~al. (2000)]{english2000} English, E., Taylor, A. R., Mashchenko, S. Y., Irwin , J. A., Basu, S., \& Johnstone, D. 2000, \apj, 533, L25
\bibitem[Freyer et~al. (2003)]{freyer2003} Freyer, T., Hensler, G,, \& Yorke, H. W. 2003, \apj, 594, 888
\bibitem[Garc\'{i}a-Segura \& Mac Low (1995)]{garcia1995} Garc\'{i}a-Segura, G., \& Mac Low, M.-M. 1995, \apj, 455, 160
\bibitem[Garc\'{i}a-Segura et~al. (1996)]{garcia1996} Garc\'{i}a-Segura, G., Mac Low, M.-M., \& Langer, N. 1996, \aap, 305, 229
\bibitem[Heiles (1979)]{heiles1979} Heiles, C. 1979, \apj, 229, 533
\bibitem[Heiles (1984)]{heiles1984} Heiles, C. 1984, \apjs, 55, 585
\bibitem[Higgs et~al. (1994)]{higgs1994} Higgs, L.A., Wendker, H. J., \& Landecker, T. L. 1994, \aap, 291, 295
\bibitem[Howarth \& Prinja (1989)]{howarth1989} Howarth, I. D., \& Prinja, R. K. 1989, \apjs, 69, 527
\bibitem[Mezger \& Henderson (1967)]{mezger1967} Mezger, P. G., \& Henderson, A. P. 1967, \apj, 147, 471
\bibitem[McClure-Griffiths et~al. (2000)]{mcclure2000} McClure-Griffiths, N. M., Dickey, J. M., Gaensler, B. M., Green, A. J., Haynes, \& R. F., Wieringa, M. H. 2000, \aj, 119, 2828
\bibitem[McClure-Griffiths et~al. (2001)]{mcclure2001} McClure-Griffiths, N. M., Green, A. J., Dickey, J. M., Gaensler, B. M., Haynes, \& R. F., Wieringa, M. H. 2001, \apj, 551, 394
\bibitem[McClure-Griffiths et~al. (2002)]{mcclure2002} McClure-Griffiths, N. M., Dickey, John M., Gaensler, B. M., \& Green, A. J. 2002, \apj, 578, 176
\bibitem[Normandeau et~al. (1996)]{normandeau1996} Normandeau, M., Taylor, A. R., \& Dewdney, P. E. 1996, \nat, 380, 687
\bibitem[Normandeau et~al. (2000)]{normandeau2000} Normandeau, M., Taylor, A. R., Dewdney, P. E., \& Basu, S. 2000, \aj, 119, 2982
\bibitem[Price et~al. (2001)]{price2001} Price, S. D., Egan, M. P., Carey, S. J., Mizuno, D. R., \& Kuchar, T. A. 2001, \aj, 121, 2819 
\bibitem[Rubin (1968)]{rubin1968} Rubin, R. H. 1968, \apj, 154, 391
\bibitem[Reich \& Reich (1986)]{reich1986} Reich, P.,\& Reich, W. 1986, A\&AS, 63, 205
\bibitem[Reich et~al. (1990)]{reich1990} Reich, W., Reich, P., \& F\"urst 1990, A\&AS, 83, 539
\bibitem[R\`{o}\.{z}yczka \& Tenorio-Tagle (1985)]{rozyczka1985} R\`{o}\.{z}yczka, M., \& Tenorio-Tagle, G. 1985, \aap, 147, 202
\bibitem[Sault et~al. (1996)]{sault1996} Sault, R. J., Staveley-Smith, L., \& Brouw, W. N. 1996, A\&AS, 120, 375    
\bibitem[Stil \& Irwin (2001)]{stil2001} Stil, J. M., \& Irwin, J. A. 2001, \apj, 563, 816
\bibitem[Taylor et al.(2002)]{taylor2002} Taylor, A. R., Stil, J. M., Dickey, J. M., McClure-Griffiths, N. M., Martin, P. G., Rothwell, T. A., \& Lockman, F. J. 2002, in ASP Conf. Ser. 276, Seeing Through The Dust: The Detection Of \HI And The Exploration Of The ISM In Galaxies, ed. A. R. Taylor, T. L. Landecker, \& A. G. Wills (San Francisco: ASP), 68
\bibitem[Taylor et al.(2003)]{taylor2003} Taylor, A. R., Gibson, S. J., Peracaula, M., Martin, P. G., Landecker, T. L., Brunt, C. M., Dewdney, P. E., Dougherty, S. M., Gray, A. D., Higgs, L. A., Kerton, C. R., Knee, L. B. G., Kothes, R., Purton, C. R., Uyaniker, B., Wallace, B. J., Willis, A. G., \& Durand, D. 2003, \aj, 125, 3145 
\bibitem[Treffers \& Chu (1982)]{treffers1982} Treffers, R. R., \& Chu, Y.-H. 1982, \apj, 254, 569
\bibitem[Uyaniker \& Kothes(2002)]{uyaniker2002} Uyaniker, B., \& Kotes, R. 2002, \apj, 574, 805
\bibitem[Vacca et~al. (1996)]{vacca1996} Vacca, W. D., Garmany, C. D., \& Shull, J. M. 1996, \apj, 460, 914
\bibitem[Van Buren (1986)]{vburen1986} Van Buren, D. 1986, \apj, 306, 538
\bibitem[Van der Werf \& Higgs (1990)]{vdwerf1990} Van der Werf, P. P., \& Higgs L. A. 1990, \aap, 235, 407
\bibitem[Weaver et~al. (1977)]{weaver1977} Weaver R., McCray R., \& Castor, J. 1977, \apj, 218, 377
\end{thebibliography}
\end{document}